\documentclass[reprint,twocolumn,superscriptaddress,secnumarabic,amssymb,nobibnotes,aps,prx]{revtex4-2}
\usepackage{graphicx}
\usepackage{dcolumn}
\usepackage{bm}
\usepackage{xcolor}
\usepackage[export]{adjustbox}
\usepackage{float}
\usepackage{tabularx}
\usepackage{multirow}
\usepackage{amsmath}
\usepackage{braket}
\usepackage{array}
\usepackage{ulem}
\usepackage{comment} 

\newcolumntype{C}[1]{>{\centering\let\newline\\\arraybackslash\hspace{0pt}}m{#1}}

\begin{document}
	
	
	\title{UTe$_2$: a narrow band superconductor}
	
	\author{Martin Sundermann}
	\altaffiliation{M.S. and T.O. contributed equally to this work.}
	\affiliation{Max Planck Institute for Chemical Physics of Solids, N{\"o}thnitzer Stra{\ss}e 40, 01187 Dresden, Germany}
	\affiliation{PETRA III, Deutsches Elektronen-Synchrotron DESY, Notkestra{\ss}e 85, 22607 Hamburg, Germany}
	
	\author{Takaki Okauchi}
	\altaffiliation{M.S. and T.O. contributed equally to this work.}
	\affiliation{Department of Physics and Electronics, Osaka Metropolitan University 1-1 Gakuen-cho, Nakaku, Sakai, Osaka 599-8531, Japan} 
	
	\author{Naoki Ito}
	\affiliation{Department of Physics and Electronics, Osaka Metropolitan University 1-1 Gakuen-cho, Nakaku, Sakai, Osaka 599-8531, Japan} 
	
	\author{Denise~S.~Christovam}
	\altaffiliation{present address: Max Planck Institute for Solid State Research, Heisenbergstra{\ss}e 1, 70569 Stuttgart,  Germany}
	\affiliation{Max Planck Institute for Chemical Physics of Solids, N{\"o}thnitzer Stra{\ss}e 40, 01187 Dresden, Germany}
	
	\author{Andrea~Marino}
	\affiliation{Max Planck Institute for Chemical Physics of Solids, N{\"o}thnitzer Stra{\ss}e 40, 01187 Dresden, Germany}
	
	\author{Daisuke~Takegami}
	\affiliation{Max Planck Institute for Chemical Physics of Solids, N{\"o}thnitzer Stra{\ss}e 40, 01187 Dresden, Germany}
	\altaffiliation{Present address: Department of Applied Physics, Waseda University, Shinjuku, Tokyo 169-8555, Japan}
	
	\author{Andrei Gloskovskii}
	\affiliation{PETRA III, Deutsches Elektronen-Synchrotron DESY, Notkestra{\ss}e 85, 22607 Hamburg, Germany}
	
	\author{Priscila~F.~S.~Rosa}
	\affiliation{Los Alamos National Laboratory, Los Alamos, New Mexico 87545, USA}
	
	\author{Jan~Kune{\v s}}
	\affiliation{Department of Condensed Matter Physics, Faculty of Science, Masaryk University, Kotl\'a\v{r}sk\'a 2, 611 37 Brno, Czechia}
	
	\author{Shin-ichi~Fujimori}
	\affiliation{Materials Sciences Research Center, Japan Atomic Energy Agency, Sayo, Hyogo 679-5148, Japan}
	
	\author{Liu~Hao~Tjeng}
	\affiliation{Max Planck Institute for Chemical Physics of Solids, N{\"o}thnitzer Stra{\ss}e 40, 01187 Dresden, Germany}
	
	\author{Andrea~Severing}
	\altaffiliation{corresponding author: andrea.severing@cpfs.mpg.de}
	\affiliation{Max Planck Institute for Chemical Physics of Solids, N{\"o}thnitzer Stra{\ss}e 40, 01187 Dresden, Germany}
	\affiliation{Institute of Physics II, University of Cologne, Z\"{u}lpicher Stra{\ss}e 77, 50937 Cologne, Germany}

	\author{Atsushi~Hariki}
	\altaffiliation{corresponding author: hariki@omu.ac.jp}
	\affiliation{Department of Physics and Electronics, Osaka Metropolitan University 1-1 Gakuen-cho, Nakaku, Sakai, Osaka 599-8531, Japan}
	\date{\today}

	\begin{abstract}
		We investigate the nature of the 5$f$ electrons in the unconventional odd-parity superconductor UTe$_2$, focusing on the degree of covalency, localization versus itinerancy, and dominant electronic configuration. This is achieved using density functional theory (DFT) in combination with dynamical mean-field theory (DMFT) calculations. A key aspect of our approach is the material-specific tuning of the double-counting correction parameter, $\mu_{\rm dc}$, within the DFT\,+\,DMFT part. This tuning is guided by the energy dependence of photo-ionization cross-sections in valence band photoelectron spectroscopy. The reliability of the parameters is confirmed by the accurate reproduction of the angle-resolved valence-band photoemission spectra and the U\,4$f$ core-level data. The DFT\,+\,DMFT model reveals that in UTe$_2$ U\,5$f^n$ configurations with \textit{n}\,=\,1 to 4 contribute to the ground state, with the 5$f^2$ configuration being most prevalent and an average 5$f$ shell fillings close to 2.5. The model further suggests that the 5$f$ electrons form narrow bands and that charge fluctuations due to degeneracy play a role in addition to coherent valence dynamics arising from hybridization with the conduction bath.  Additionally, the significance of the U\,6$d$ states in UTe$_2$ is discussed. 
	\end{abstract}

	\maketitle
	
	\section{Introduction}
	The heavy-fermion and spin-triplet superconductor UTe$_2$\,\cite{Ran2019,Aoki2019,Visser2019} has generated significant interest due to the potential for the triplet pairing to induce non-trivial topological properties. UTe$_2$ crystallizes in the orthorhombic $Immm$ structure, exhibiting distinct anisotropies in transport properties and magnetization in the normal state. Superconductivity emerges below the critical temperature $T_c$\,=\,2.1\,K~\cite{Sakai2022}, with a highly anisotropic upper critical field that far exceeds the Pauli limit expected for even-parity superconductors\,\cite{Lewin2023,Wu2024}.
	
	The phase diagram of UTe$_2$ is highly tunable, exhibiting competing superconducting and (meta)magnetic phases. Two superconducting phases are observed as a function of applied \textit{b}-axis field, with a metamagnetic transition to a field-polarized phase at even higher fields, which includes a field-induced superconducting pocket for fields applied 35$^{\circ}$ away from the \textit{b} axis. Moderate hydrostatic pressure initially suppresses superconductivity but later another superconducting phase appears\,\cite{Knebel2019,Ran2019b,Lewin2023}. As pressure increases further, an antiferromagnetic phase emerges\,\cite{Knafo2025,Thomas2020}, followed by a phase transition from orthorhombic to tetragonal ($I4/mmm$) structure between around 5 and 7\,GPa\,\cite{Ran2020,Thomas2020,Huston2022,Honda2023,Deng2024}. This transition is marked by a volume collapse of about 10\% and a 4\% increase of the uranium-uranium distance $d_{\rm UU}$. Notably, $d_{\rm UU}$ is above the Hill limit at all pressures. Applied pressure also affects the average orbital moment as well as the average 5$f$-shell filling, $\langle n_f\rangle$, which initially decreases but recovers at the crystallographic phase transition\,\cite{Wilhelm2023,Deng2024}.

	In this work, we address the degree of covalence of the 5$f$ electrons in UTe$_2$, as the hybridization of $f$ electrons with the conduction electron bath plays a crucial role in the physics of actinide compounds. A key challenge in uranium compounds is the strong hybridization between the 5$f$ and conduction electrons, due to the spatially extended nature of the 5$f$ shell. Additionally, multiple electronic configurations contribute to the ground state, resulting in the average uranium valence deviating strongly from integer values\,\cite{Allen1987,Kotani1992}.	Consequently, determining the most relevant electron configuration, typically 5$f^2$ or 5$f^3$, that defines the ground state’s quantum numbers, as well as understanding the distribution of configurations contributing to the ground state, provides important insights and a route towards the long-sought microscopic model of actinide materials.
	
	Unfortunately, in uranium compounds this information remains elusive in commonly applied experiments. X-ray absorption (XAS) and core-level photoelectron spectroscopy (PES) data are broad and structureless and do not provide sufficient information for a quantitative analysis of the many-body ground state\,\cite{Groot2008}. 
	This challenge, along with the necessity of an Anderson-impurity-model (AIM) based approach accounting for both the strong correlation of U 5$f$ electrons and their hybridization, has resulted in conflicting interpretations of uranium spectroscopy data and ongoing discussions about the 
	dominant local electronic configuration in the normal state\,\cite{Fujimori2019,Miao2020,Thomas2020,Fujimori2021,Shick2021,Aoki2022b,Liu2022,Wilhelm2023}. 
	
	Recently, some of us demonstrated how the dilemma regarding the ground state configuration can be addressed using valence band resonant inelastic x-ray scattering (VB-RIXS). VB-RIXS offers unique insights into the dominant electronic configuration, particularly when performed at the U\,$M_{4,5}$ edges, where the signal-to-background ratio is excellent\,\cite{Marino2023}. Applying this technique to UTe$_2$, multiplet excitations characteristic of the 5$f^2$ configuration\,\cite{Christovam2024} were observed. The detection of these excitations, along with successful modeling using ionic calculations, suggests that the 5$f$ electrons in UTe$_2$ are strongly correlated \cite{Liu2022, Christovam2024}. Additionally, the observation of charge transfer scattering in $M$-edge VB-RIXS spectra provides further evidence of the strongly non-integer valent nature of UTe$_2$.
	
	This raises the question of the \textit{degree} of correlation or itinerancy in the non-integer valent ground state of UTe$_2$. In principle, photoelectron spectroscopy (PES) is sensitive to covalency\,\cite{Gunnarsson1983,Gunnarsson1988,Kotani1988,Kotani1999}, but intrinsic broadening also limits the information of uranium core-level PES spectra; they exhibit a single broad emission line and at most one satellite feature, see, e.g., Refs.~\cite{Fujimori2012,Fujimori2016,Fujimori2021}.

	In Ref.\,\cite{Marino2024}, we demonstrated that interpreting U\,4$f$ core-level data for uranium intermetallic compounds is not straightforward. Using two isostructural model compounds, Pauli paramagnetic UB$_2$ and ferromagnetic UGa$_2$ ($T_C$\,=\,125\,K, $\mu_{\rm ord}$\,=\,3\,$\mu_B$), which represent the more extreme ends of the de-/localization spectrum, we showed that density functional theory (DFT) combined with dynamical mean field theory (DMFT) with material specific parameters, is essential to account for correlation effects and to gain insight the degree of de/localization. Key computational parameters, including the double counting correction $\mu_{\rm dc}$, the Hubbard $U_{\rm ff}$ and Hund's $J$ of the U\,$5f$ states, were optimized to accurately reproduce the valence band (VB) spectra measured with PES using both hard and soft x-rays. The use of two very different x-ray energies allows for clear distinction between the correlated 5$f$ and the other, weakly correlated, orbitals 	near the Fermi energy, due to the different energy dependence of their respective photoionization cross-sections~\cite{TRZHASKOVSKAYA2001,TRZHASKOVSKAYA2002,TRZHASKOVSKAYA2018}: In the soft x-rays regime, the signal from the correlated 5$f$ states dominates, while at 6000\,eV, the non-5$f$ states are strongly enhanced.
	
	In this study, we extend this approach to UTe$_2$, utilizing DFT\,+\,DMFT calculations with parameters tuned to reproduce experimental VB PES spectra at both 6000 and 800\,eV, angle resolved PES (ARPES) at 565--675\,eV as well as core-level PES. Once the electronic model is established by setting the parameters ($\mu_{\rm dc}$, $U_{\rm ff}$, and $J$) in the model Hamiltonian the respective weights of 5$f$ configurations in the intermediate valent ground state are determined and the time dependent charge correlation function is calculated. 
	
	\section{Methods}
	\subsection{Experiment}
	The VB PES and ARPES data, measured at beamline BL23SU at SPring-8\,\cite{Saitoh2012} at 20\,K with 800\,eV and 556-675\,eV incident energy, respectivey, and energy resolutions of 140\,meV and 100\,meV, are adapted from Ref.\,\cite{Fujimori2019}. The hard x-ray VB PES data with 6000\,eV incident energy and resolution of 230\,meV are adapted from \,\cite{Christovam2024}. The core-level, PES measurements with hard x-rays (HAXPES) were performed on the same set of plate-like single crystalline samples of UTe$_2$ as in Ref.\,\cite{Christovam2024} and measured under the same conditions, i.e., at P22 beamline\,\cite{schlueter2019} at PETRA III (DESY) in Hamburg, Germany with 6000\,eV incident energy, an overall resolution of $\approx230$\,meV and temperature $T$\,=\,40\,K. 
	
	\subsection{Calculations}
	The computational modeling of UTe$_2$ followed the same philosophy as in Ref.~\cite{Marino2024}. We began by performing a standard DFT calculation for the experimental crystal structure~\cite{Hutanu2020}, using the local density approximation for the exchange-correlation functional and incorporating spin-orbit coupling, as implemented in the \textsc{Wien2k} package~\cite{wien2k,sm}. Based on the resulting DFT band structure, we constructed a tight-binding model including U 5$f$, 7$s$, 6$d$, 7$p$ and Te 5$s$, 5$p$ orbitals~\cite{wien2wannier,wannier90}. The lattice model was then augmented with local Coulomb interactions among the U 5$f$ electrons, parameterized by the Hubbard $U_{\rm ff}$ and Hund’s $J$, and solved using DMFT~\cite{georges96,kotliar06}. The self-energy of the U 5$f$ electrons was computed from the AIM using the continuous-time quantum Monte Carlo method~\cite{werner06,gull11,boehnke11,hafermann12}, retaining only the density-density terms of the Coulomb vertex for computational efficiency. Upon convergence of the DMFT self-consistency loop, the VB spectrum was obtained by analytic continuation of the self-energy using the maximum entropy method~\cite{jarrell96}. All calculations were performed at $T = 300$~K.
	
	In the DFT\,+\,DMFT modeling, determining the double-counting correction $\mu_{\rm dc}$ is a crucial step to account for the $f$--$f$ interaction effects already present at the DFT level~\cite{kotliar06,Karolak10,Haule15}. The U 5$f$ orbital energies in the DMFT model are obtained by shifting their corresponding DFT values by $\mu_{\rm dc}$; hence the choice of $\mu_{\rm dc}$ is essential for correctly representing the uranium valency. Although no universally accepted expression for $\mu_{\rm dc}$ exists, it effectively renormalizes the energy separations between the correlated U\,5$f$ states and the uncorrelated bands. A realistic value of $\mu_{\rm dc}$ can therefore be determined by tuning it to reproduce the experimental VB PES spectrum~\cite{Marino2024}, measured at very different photon energies as described above. To further validate the optimized DMFT model, both ARPES and core-level spectra are calculated and compared to experimental data; The ARPES comparison confirms the model's accuracy with respect to the low-energy bands, while the core-level comparison, it being sensitive to covalency and $f$-configuration energies~\cite{Marino2024}, offers complementary support for the electronic model. The computational details for these spectra follow the methodology of Refs.~\onlinecite{Marino2024,Hariki17}. 
	
	\begin{figure}[t]
		\begin{center}
			\includegraphics[width=0.99\columnwidth]{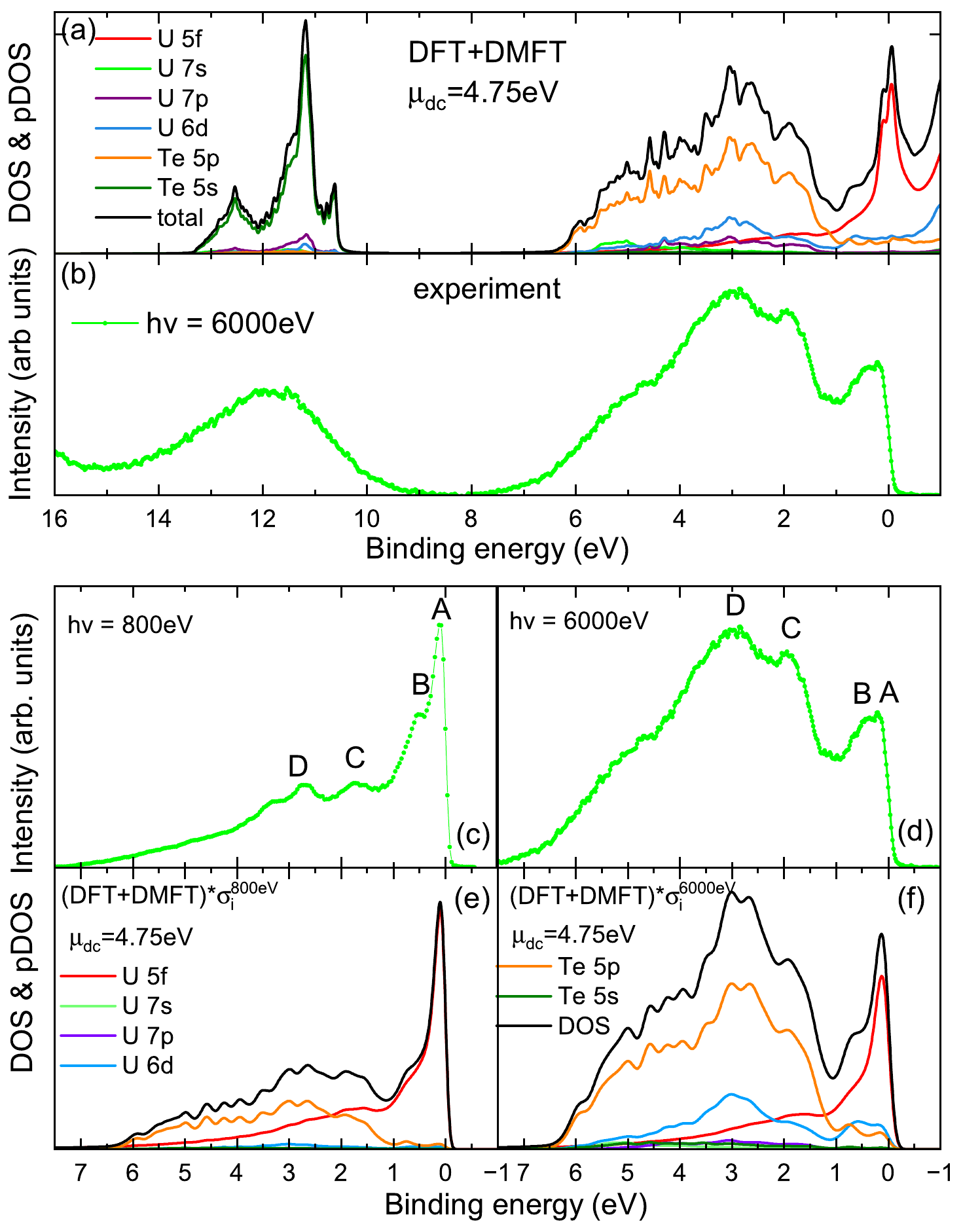}
		\end{center}
		\caption{(a) Valence band (VB) spectra: total DOS (black) and orbital-resolved DOS (color) from the DFT\,+\,DMFT calculation with $U_{\rm ff} = 3$\,eV, $J = 0.59$\,eV, and optimized $\mu_{\rm dc} = 4.75$\,eV. (b) Hard x-ray VB PES data adapted from Ref.~\cite{Christovam2024}. (c), (d) VB PES data over a smaller energy range, measured with soft~\cite{Fujimori2019} and hard x-rays, respectively. The labels $A$, $B$, $C$, and $D$ indicate characteristic spectral features. (e), (f) Total and orbital-resolved DOS from the same DFT\,+\,DMFT calculations as in (a), multiplied by the Fermi function and respective photoemission cross-sections, and broadened to account for experimental resolution.}
		\label{VB_DMFT}
	\end{figure}
	
	\begin{figure}[t]
		\begin{center}
			\includegraphics[width=0.99\columnwidth]{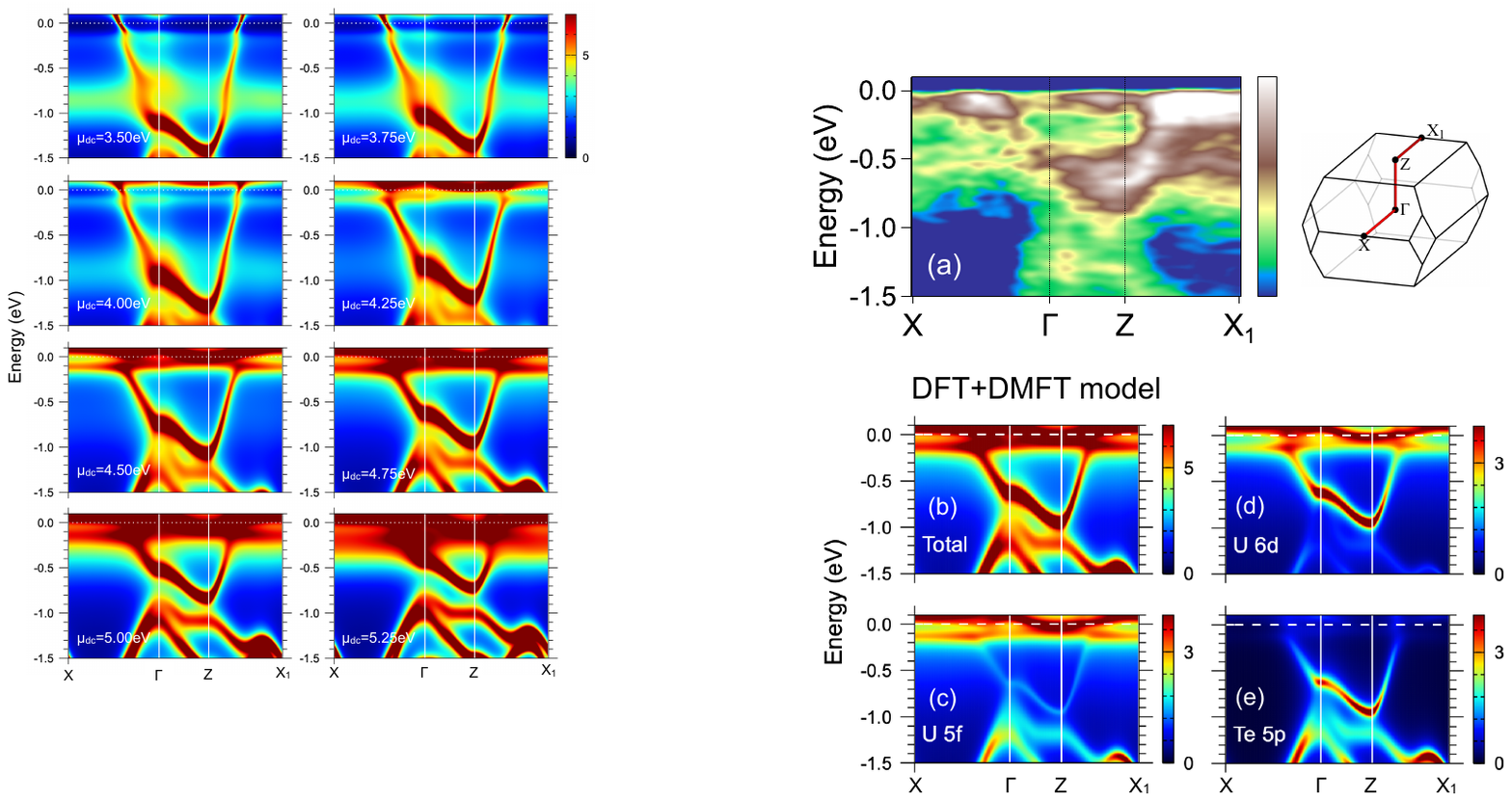}
		\end{center}
		\caption{(a) Experimental soft x-ray (800\,eV) ARPES data of UTe$_2$ along the momentum path shown in the right panel. (bottom) momentum-resolved spectral densities calculated from the optimized DFT\,+\,DMFT model. Panel (b) displays the total spectral densities, while panels (c), (d), and (e) show the orbital contributions from U 5$f$, U 6$d$, and Te 5$p$ states, respectively.}
		\label{ARPES}
	\end{figure}
	\begin{figure}[t]
		\begin{center}
			\includegraphics[width=0.98\columnwidth]{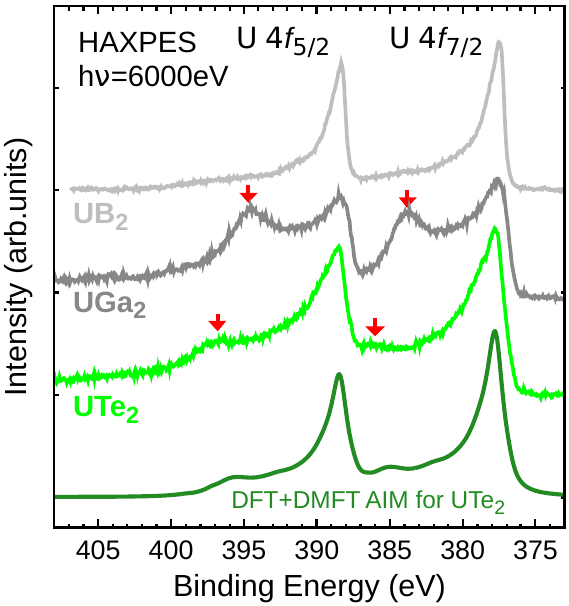}
		\end{center}
		\caption{U\,4$f$ core-level spectra of UTe$_2$ (light green), corrected for integrated background and plasmon contributions (see Appendix\,\textbf{B}), compared to those of UGa$_2$ (dark grey) and UB$_2$ (light grey), adapted from Ref.\,\cite{Marino2024}. The dark green curve shows the calculated core-level spectra of UTe$_2$ using the DFT\,+\,DMFT AIM.}
	\label{core}
\end{figure}

\section{Results and Discussion}

\subsection{DFT\,+\,DMFT modeling of UTe$_2$}

Figure\,\ref{VB_DMFT}\,(a) shows the spectral functions for UTe$_2$ as obtained from our DFT\,+\,DMFT calculations for $U_{\rm ff}$\,=\,3\,eV, $J$\,=\,0.59\,eV and the optimized value for $\mu_{\rm dc}$\,=\,4.75\,eV in the energy window -1 to 16\,eV. The comparison to experimental VB data in panel (b), here measured with hard x-rays, confirms that the main features are well captured. That is, the peak close to the Fermi energy, a broad hump-like signal between 1 and 6\,eV, and the signal close to 12\,eV. The latter is due to Te\,5$s$ states.

For further identification of subshells, especially close to the Fermi energy, we inspect the energy dependence of spectral features in a smaller energy range. In Figs\,\ref{VB_DMFT}\,(c) and (d) we compare the VB data of UTe$_2$ measured with 800\,eV and 6000\,eV incident energy.  We observe a strong change in spectral shape which is due to the different energy dependence of the subshells' cross-sections.  Most strikingly, the soft x-ray data peak sharply close to the Fermi energy (see peak labeled \textit{A}),  followed by a shoulder ($B$), and some smaller spectral weight with some structure, labeled with $C$ and $D$. Spectral weight $A$ is strongly suppressed in the hard x-ray spectra in panel (d). Only a broader hump consisting of $A$ and $B$ survives, while the broad spectral weight with features $C$ and $D$ around 2 and 3\,eV binding energy is strongly enhanced.  The 5$f$ signal being 10 times stronger than the non-5$f$ in the 800\,eV data, lets us safely conclude the feature closest to the Fermi energy must be mainly due to the 5$f$ states. In contrast, features $C$ and $D$ must consist mainly of non-5$f$ states.  

Figures\,\ref{VB_DMFT}\,(e) and (f) show the same DFT\,+\,DMFT VB spectra as in panel (a), now multiplied with the Fermi function and broadened to account for resolution effects, and corrected for the respective cross-sections for 800 and 6000\,eV (see Table\,I Appendix\,\textbf{A}). A comparison between the experimental data in (c) and (d) and total calculated density of states (DOS) (black lines) in (e) and (f) demonstrates that the intensity variations of spectral weights close to the Fermi energy, as well as the broad distribution around 3\,eV with varying incident energy are best reproduced for the present choice of parameters. In Appendix\,\textbf{C}, Fig.\,\ref{LDA_mu}\,(a), we show the sensitivity of the calculated VB spectra to $\mu_{\rm dc}$.

Upon close inspection of Fig.\,\ref{VB_DMFT}\,(c)-(f), we find that the intensities $A$ and $B$ near the Fermi energy are predominantly due to the U\,5$f$ states (red lines), but not purely. Also Te\,5$p$ (orange) and U\,6$d$ (blue) states have some spectral weight below 1\,eV binding energy. Meanwhile, the spectral feature $C$ and $D$ are primarily composed of Te\,5$p$ states, with some contribution from the U\,5$f$ and U\,6$d$ states. 

To further validate our choice of parameters, experimental ARPES spectra are compared with the optimized DFT\,+\,DMFT results along selected momentum paths, see Fig.\,\ref{ARPES}. We find good agreement between experiment and theory when keeping in mind that the soft x-ray experiment enhances the 5$f$ states with respect to the other states, something not accounted for in the momentum resolved spectral functions of Figs.\,\ref{ARPES}: The flat branch close to the Fermi energy and also the strongly dispersive band along the $\Gamma$--$Z$--$X_1$ path down to 1\,eV are reproduced. The orbital-resolved spectral contributions from U\,5$f$, U\,6$d$, and Te\,5$p$ states, shown in Figs.\,\ref{ARPES}(c)--(e), reveal that the strongly dispersive band has dominant U\,6$d$ and Te\,5$p$ character. In addition, the narrow band near the Fermi level, as well as the enhanced spectral weight around 0.3--0.5\,eV near the $X_1$ point is due to 5$f$ states. This agreement deteriorates rapidly when the double-counting parameter, i.e., the position of the U\,5$f$ level, is shifted away from its optimal value (see Appendix \textbf{C}, Fig.\,\ref{ARPES_mu}).

The model derived above also reproduces the U\,4$f$ core-level spectrum of UTe$_2$, as shown in Fig.\,\ref{core} (see green dots), alongside the spectra of the two model compounds from Ref.\,\cite{Marino2024} (see grey dots). The present hard x-ray data of UTe$_2$ are consistent with previous soft x-ray results reported in Refs.\,\cite{Fujimori2016,Fujimori2019a,Fujimori2021}. The core-level spectra in Fig.\,\ref{core} exhibit strong material dependence, with both the spectral weight and energy position of the satellite relative to the main line varying significantly (see red arrows). We calculated the core-level spectrum using the optimized DMFT model, incorporating the core-valeance interaction $U_{\rm fc}$ in the DFT\,+\,DMFT impurity framework. Good agreement with experiment is obtained for $U_{\rm fc} = 5.0$\,eV (see dark green line), a value consistent with those used for UGa$_2$ and UB$_2$\,\cite{Marino2024}. In Appendix\,\textbf{C}, Fig.\,\ref{corelevel_mu} we show the sensitivity of the core-level line shape to $\mu_{\rm dc}$ and $U_{\rm fc}$, respectively, confirming that agreement with experiment is achieved only near this optimal $\mu_{\rm dc}$ value.

\begin{figure*}[t]
	\begin{center}
		\includegraphics[width=1.99\columnwidth]{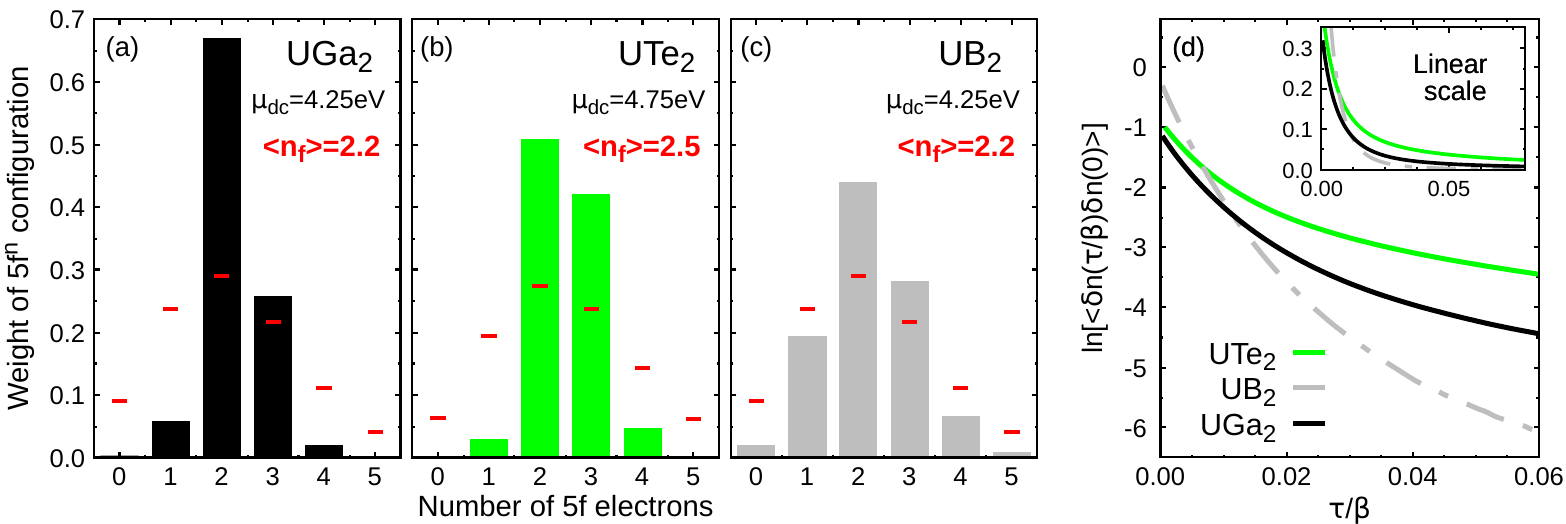}
	\end{center}
	\caption{(a)-(c): Valence histograms from DFT\,+\,DMFT of UGa$_2$, UTe$_2$ and UB$_2$, respectively, and binomial distribution (red ticks). (d) Logarithm of charge correlation function $\langle \delta n(\tau) \delta n(0)\rangle$ from DFT\,+\,DMFT for of UTe$_2$ (green) and the two model compounds UGa$_2$ (black) and UB$_2$ (light grey). The data of UGa$_2$ and UB$_2$ are adapted from Ref.\,\cite{Marino2024}. A linear-scale plot is shown in the inset.}  
	\label{hist}
\end{figure*}

\subsection{U 5$f$ occupation and charge correlation function}

Having established the electronic model that reproduces the VB, ARPES and core-level spectra, we now use this model to examine the degree of correlation and covalency in UTe$_2$. In Fig.\,\ref{hist}, we start with the U valence histogram, computed from the DFT\,+\,DMFT results using the optimized model parameters, and compare it to the respective histograms of the model compounds UGa$_2$ and UB$_2$ from Ref.\,\cite{Marino2024}. Additionally, we overlay the binomial distributions corresponding to the fully itinerant, non-interacting band-like limit for the average filling given by DFT\,+\,DMFT\,\cite{phdthesis_bosch}. For completeness, the variation of the 5$f^n$ spectral weights with $\mu_{\rm dc}$ is shown in Appendix\,\textbf{C}, Fig.\ref{LDA_mu}\,(b).

In Ref.\,\cite{Marino2024}, we showed, that UGa$_2$ is a strongly correlated compound dominated by the 5$f^2$ configuration, while UB$_2$ represents a strongly itinerant system, close to the band-like limit. Here, the width of the distribution of 5$f^n$ configurations in the histogram serves as a measure of the magnitude of charge fluctuations between the relevant 5$f$ configurations.

In UTe$_2$, we also find that the 5$f^2$ configuration is the most prominent, consistent with the 5$f^2$ multiplet structure observed in the VB-RIXS spectra\,\cite{Liu2022,Christovam2024}. Moreover, the distribution of 5$f^n$ configurations deviates significantly from a binomial distribution and is considerably narrower than that of UB$_2$, reinforcing the characterization of UTe$_2$ as a strongly correlated material. However, the distribution in UTe$_2$ is slightly wider than that of UGa$_2$, indicating that the degree of charge dynamics in UTe$_2$ differs from that of the prototypically correlated uranium compound UGa$_2$. In fact, the histogram in UTe$_2$ is mainly composed of the configurations $5f^2$ and $5f^3$, with the center of gravity shifted towards higher occupations compared to UGa$_2$ and UB$_2$, resulting in an average 5$f$-shell occupancy of $\langle n_{\rm f}\rangle $\,=\,2.5. 

Further insight into the differences of the three compounds can be gained by examining the time dependence of the local charge-charge correlation function $\langle \delta n(\tau) \delta n(0)\rangle$ in the DFT+DMFT results. In Fig.\,\ref{hist}\,(d), we plot the logarithm of this function in the imaginary time domain for UTe$_2$ (green line) and, for comparison, also for UGa$_2$ and UB$_2$~\cite{Marino2024} (black and gray lines). 

The instantaneous value at $\tau = 0$ reflects the magnitude of charge fluctuations. UTe$_2$ exhibits stronger charge fluctuations than UGa$_2$, but weaker than UB$_2$, consistent with the respective width of the valence histograms in Figs.\,\ref{hist}\,(a)--(c).	At finite $\tau$, the charge correlation function of UTe$_2$ decays more slowly than that of UGa$_2$, and exhibits a more pronounced constant component. Charge fluctuations between configurations may arise from two distinct mechanisms: 
a) Covalency and itinerancy  lead to a coherent admixture of valence states and is referred to as \textit{intermediate valency}. It results in a rapid decay of $\langle \delta n(\tau) \delta n(0)\rangle$, as clearly observed in UB$_2$ and, to a lesser extent, in UGa$_2$. 
b) Quasi-degenerate configurations, such as, e.g. 5$f^2$ and 5$f^3$, which give rise to long-lived, local fluctuations between valence states, are referred to as \textit{mixed valency}\,\cite{Ylvisaker2009}. In the extreme case of fully degenerate configurations without intersite coherence, this would produce a constant value of $\langle \delta n(\tau) \delta n(0)\rangle$ at finite $\tau$. In UTe$_2$, the value at $\tau$\,=\,0 indicates stronger charge fluctuations than in UnGa$_2$, while the slower decay at finite $\tau$ reflects a different balance between the two charge-fluctuation mechanisms in these two materials. In UTe$_2$, the charge fluctuations stem to a larger extend from the degeneracy of 5\textit{f} configurations, that is, from a smaller energy separation between the 5$f^2$ and 5$f^3$ states. However, because $\langle \delta n(\tau) \delta n(0)\rangle$ does not remain strictly constant over time, hybridization with the surrounding electron bath is still present, albeit to a lesser extend than in UGa$_2$. This makes UTe$_2$ a narrower-band material compared to UGa$_2$, consistent  with the high tunability of its phase diagram\,\cite{Knebel2019,Wu2024,Ran2019b,Lewin2023,Knafo2025,Ran2020,Thomas2020,Huston2022,Honda2023,Deng2024,Wilhelm2023}.

The analysis of the charge correlation function provides insights into the satellites observed in the core-level spectra, Fig.\,\ref{core}. The intensity and energy of these satellites are determined by a delicate balance among covalency, electron correlation, and configuration mixing. This analysis suggests that the satellites originate from distinct physical mechanisms: the weak satellite in UB$_2$ is a consequence of strong covalency; the strong satellite in UGa$_2$ is due to strong  correlations and localization; while in UTe$_2$, it emerges from a subtle balance between $f^2$ and $f^3$ mixed valency.

\subsection{Choice of model parameters in DFT\,+\,DMFT}

Several DFT\,+\,DMFT studies have been performed on UTe$_2$, each adopting different choices of $U_{\rm ff}$ and $J$ parameters, and different treatments of $\mu_{\rm dc}$, leading to variations in the U\,5$f$ spectral features near the Fermi level. Most of these studies estimate the U\,5$f$ occupation close to 2, in contrast to our estimate of 2.5. For example, Refs.~\cite{Miao2020,Xu2019} adopt a large $U_{\rm ff}$\,=\,6-8\,eV together with the so-called nominal double-counting scheme for $\mu_{\rm dc}$. To our knowledge, applying the nominal $\mu_{\rm dc}$ scheme, which assumes an electron count $f^2$ (or $f^3$) when evaluating the correction parameter $\mu_{\rm dc}$, to a mixed-valent $f$-electron system such as UTe$_2$ has not been validated against experimental data. In Ref.~\cite{Choi2024}, the $U_{\rm ff}$ value is adjusted to reproduce a specific feature in the ARPES data, but neither the $\mu_{\rm dc}$ value nor the resulting U\,5$f$ occupation are provided. Ref.~\cite{Kang2024}, on the other hand, uses a different implementation of DMFT than the other works and reports a 5$f$ occupation of 2.3. 

In the following, we elaborate on our choice of model parameters for UTe$_2$ within the DFT\,+\,DMFT framework. To assess parameter sensitivity, we performed additional calculations using slightly different sets of $U_{\rm ff}$ and $J$, varying $J$ between 0.47 and 0.7\,eV for a fixed $U_{\rm ff}$\,=\,2.8\,eV, with corresponding adjustments to $\mu_{\rm dc}$. While different combinations of $U_{\rm ff}$ and $J$ yield variations in both spectral shape and U\,5$f$ occupation, we find that the VB spectra converge to a similar shape that reasonably reproduces the experimental VB-PES data, provided that $\mu_{\rm dc}$ is individually tuned (Appendix\,\textbf{C}, Fig.\,\ref{parameters}). Minor differences in spectral peak positions or intensity ratios remain, as also discussed in Ref.~\cite{Marino2024}. Crucially, such agreement with experiment is achieved only when $\mu_{\rm dc}$ is adjusted such that the distribution of the 5$f^n$ configurations in the ground state and average 5$f$ occupation $\langle n_{\rm f}\rangle $ remains about the same, regardless of the specific values of $U_{\rm ff}$ and $J$ within the studied range. 

This underscores the critical role of accurately determining $\mu_{\rm dc}$ for predicting the correct uranium valency. As shown in Fig.\,\ref{LDA_mu}\,(b) of Appendix\,\textbf{C}, even small variations in $\mu_{\rm dc}$ can significantly affect the calculated 5$f$ occupation. While certain low-energy band features may be captured by adjusting $U_{\rm ff}$ or $J$ alone, such adjustments lead to either over- or underestimation of the 5$f$ occupation. This discrepancy in electron count immediately manifests in core-level PES, which is highly sensitive to the uranium valency, as demonstrated in Fig.~\ref{corelevel_mu}. Therefore, the direct comparison with the comprehensive set of spectroscopic data presented here provides a valuable benchmark for evaluating the reliability of the chosen parameters and the underlying electronic structure of UTe$_2$ with its nearly degenerate U\,5$f^2$ and 5$f^3$ configurations and narrow 5$f$ bands.

\subsection{Importance of U\,6$d$ states} 
Our optimized DFT\,+\,DMFT model raises the question of how such a complex mixed-valence state is accommodated in the crystal structure of UTe$_2$. As pointed out in Ref.\,\cite{Christovam2024}, one of the two Te sites cannot spatially accommodate a large Te$^{2-}$ ion, which lead to the proposal of a hypothetical U$^{3+}$ valence state, typically associated with a formal U\,5$f^3$ configuration. However, both the valence histogram obtained in this study and the previously reported RIXS data indicate the relevance of the 5$f^2$ configuration. To reconcile this, Miao \textit{et al.}\,\cite{Miao2020} and Christovam \textit{et al.}~\cite{Christovam2024} suggested that the U\,6$d$ orbitals, particularly the 6$d_{z^2-r^2}$ states, must absorb some of the charge. 

To test this conjecture we performed DFT\,+\,DMFT calculations with a modified model in which electron transfer between U\,5$f$ and U\,6$d$ states was effectively suppressed. This was achieved by shifting the onsite energies of the U\,6$d$ states by 3\,eV, eliminating the U\,6$d$ states from the top of the VB. In Fig.\,\ref{modified_LDA} of Appendix\,\textbf{D}, we compare the original (a) and modified (b) DFT spectra. 	Although the U\,6$d$ weight near the top of the VB is not substantial in the original DFT model, it has a sizable impact on the U\,5$f$ valence state when correlations are taken into account. In the DFT\,+\,DMFT results for the modified model, a significant increase of the 5$f$ occupation is observed. The 5$f^3$ configuration becomes the most abundant over a broad  range of the $\mu_{\rm dc}$ values, see Appendix\,\textbf{D}, Fig.\,\ref{modified_model}\,(b). This confirms the above conjecture that U\,6$d$ states provide the reservoir, which fixes the $5f$ occupation.

\section{Summary}

We investigated the degree of covalency, the nature of valence fluctuations, and the dominant 5$f$ electronic configuration in UTe$_2$. To this end, DFT\,+\,DMFT calculations, using material specific parameters derived from both hard and soft x-ray valence-band photoelectron spectroscopy data, were presented. Our DFT\,+\,DMFT model simultaneously reproduces the valence-band spectra, ARPES data, and U\,4$f$ core-level spectra, validating the chosen parameters. Our analysis underscores the importance of anchoring the choice of computational parameters, especially the double-counting correction $\mu_{\rm dc}$ within the DFT\,+\,DMFT scheme, to experimental photoemission spectra in order to obtain reliable estimates of the 5$f$ shell occupancy.

Our results yield an average 5$f$ electron count of 2.5, with the 5$f^2$ configuration being the most dominant and defining the quantum numbers of the ground state. UTe$_2$ is identified as a narrow-band material in which itinerancy still plays a role. We argue that the higher degree of localization in UTe$_2$, compared to the ferromagnet UGa$_2$, arises from the presence of quasi-degenerate U\,5$f^2$ and 5$f^3$ configurations, pointing to mixed-valence rather than intermediate-valence behavior.

Finally DFT\,+\,DMFT calculations on a modified model without U\,6$d$ states at the valence band show increased 5$f$ shell filling, supporting the view that the U ion tends toward a 3+ valence. This suggests that, under realistic conditions, part of the charge is carried by the 6$d$ states, as previously suggested in Ref.\,\cite{ Christovam2024}. The artificial removal of the U\,6$d$ states, combined with correlation effects, significantly alters the low-energy valence-band structure, underscoring the importance of including U\,6$d$ states in modeling UTe$_2$.

\begin{acknowledgements}

	A.H.~was supported by JSPS KAKENHI Grant Numbers 25K00961, 25K07211, 23H03816, 23H03817, 23K03324, and the 2025 Osaka Metropolitan University (OMU) Strategic Research Promotion Project (Young Researcher). S.-i.F. was funded by JSPS KAKENHI Grant Numbers 18K03553, 20KK0061 and 22H03874. Parts of the computations were performed at It4innovations funded by the Ministry of Education, Youth and Sports of the Czech Republic through the e-INFRA CZ (ID:90254). A.S. acknowledges support from the German Research Foundation (DFG) - grant N$^{\circ}$ 387555779. All authors acknowledge DESY (Hamburg, Germany), a member of the Helmholtz Association HGF, for the provision of experimental facilities. The soft X-ray experiment was performed under Proposal No. 2019A3811 at SPring-8 BL23SU. Work at Los Alamos National Laboratory was performed under the auspices of the U.S. Department of Energy, Office of Basic Energy Sciences, Division of Materials Science and Engineering under project “Quantum Fluctuations in Narrow-Band Systems”. 
	
\end{acknowledgements}

\section{Appendix}

\subsection{Cross-section correction}
Table\,I lists the photoionization cross-section used to account for the different photon energies when comparing the DFT\,+\,DMFT calculations with the experimental data. The values are calculated according to the expression $\sigma_i$\,$\times$\,(1+$\beta$)/$N$ where $N$ is degeneracy of the shell, and $\sigma$ and $\beta$ are taken from  Refs.~\cite{TRZHASKOVSKAYA2001,TRZHASKOVSKAYA2002,TRZHASKOVSKAYA2018}. For h$\nu$\,=\,800\,eV, the cross-sections on a logarithmic scale are linearly interpolated between the values for h$\nu$\,=500 and 1000\,eV. Since cross-sections for U\,7\textit{p} are not listed, we used the value for U\,7\textit{s}, based on the empirical observation that photoionization cross-sections per electron are approximately equal for the corresponding n\textit{s} and n\textit{p} shells (where n is the principal quantum number).

\begin{table}[h]
	\vspace{5pt} 
	\caption{Photoionization cross-sections in kbarn ($10^{-25}$\,m$^2$) at 800 and 6000\,eV used to scale the Te and U valence states.}
	\label{tab_S1}
	\centering
	\renewcommand{\arraystretch}{1.5} 
	\setlength{\tabcolsep}{15pt} 
	\vspace{5pt}
	\begin{tabular}{c|cc}
		state & 800\,eV & 6000\,eV \\
		\hline \hline
		U\,5$f$   & 23.7 & 0.075 \\
		U\,6$d$   & 1.75 & 0.070 \\
		U\,7$s$   & 0.84 & 0.025 \\
		U\,7$p$   & 0.84 & 0.025 \\
		Te\,5$p$  & 5.08 & 0.066 \\
		Te\,5$s$  & 7.49 & 0.143 \\
	\end{tabular}
\end{table}

\subsection{Background and plasmon correction of core-level}

\begin{figure}[]
	\begin{center}
		\includegraphics[width=0.99\columnwidth]{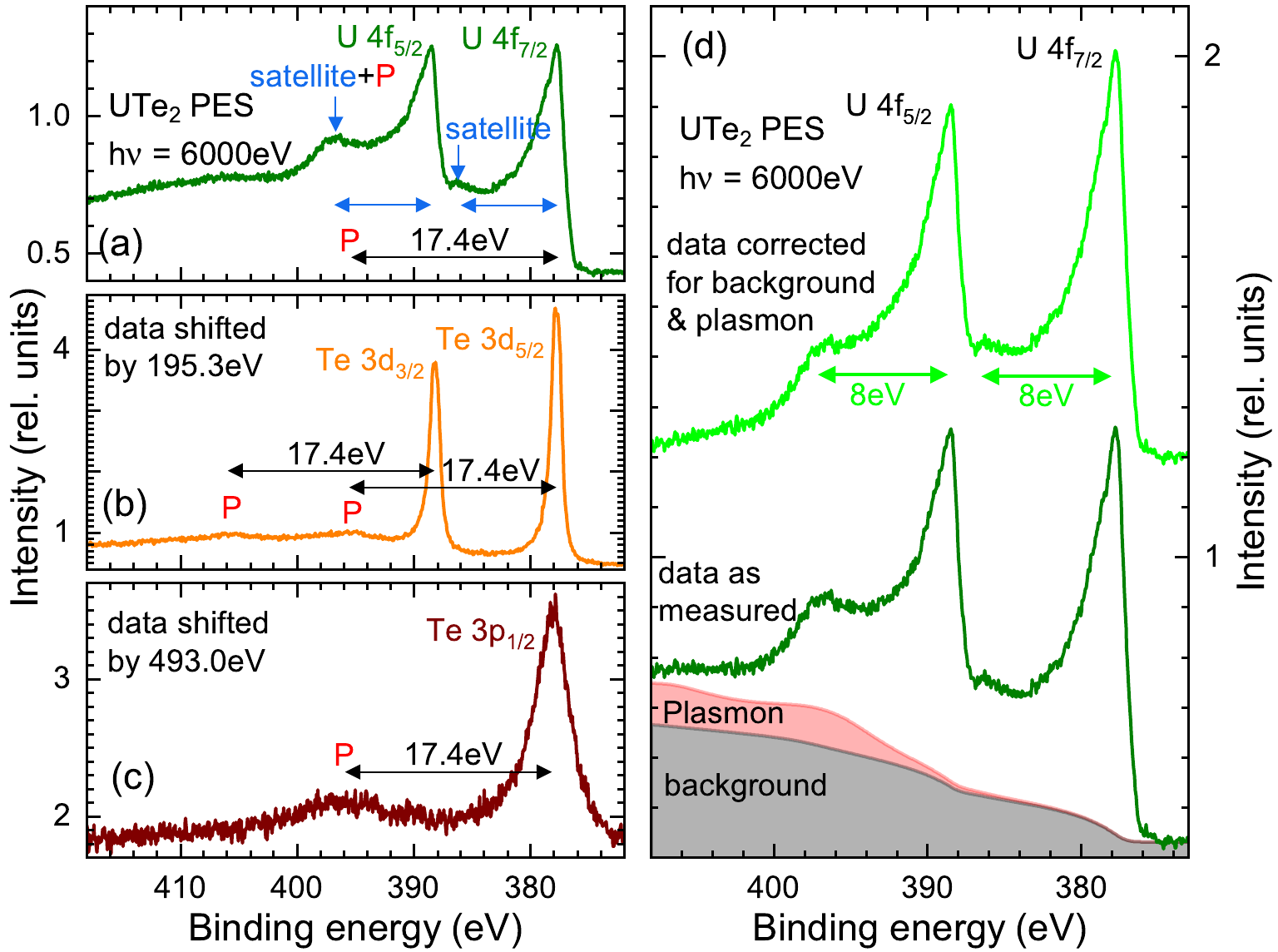}
	\end{center}
	\caption{UTe$_2$ HAXPES data: (a) U\,4$f$ core-level, (b) and (c) shifted core-level spectra of Te\,3$d$ and Te\,3$p_{3/2}$, respectively. P marks plasmon scattering at 17.4\,eV above the main absorption lines. (d) U\,4$f$ core level data of UTe$_2$ as measured (dark green) and after correction of the integral-type background and plasmon (light green) with satellites 8.2eV above the main emission lines. The filled areas below the original data represent the contribution from the integrated background (gray) and plasmons (red).}  
	\label{corelevel}
\end{figure}

Figure\,\ref{corelevel} shows U\,4$f$, Te\,3$d$ and Te\,3$p_{3/2}$ hard-ray photoelectron spectroscopy data of UTe$_2$. The Te core-level, that are not affected by configuration interaction effects, are used to correct for some minor plasmon scattering at 17.4\,eV higher binding energies from the main emission lines. The plasmons are present for all core-level. After subtracting the plasmon scattering from the U\,4$f$ core-level data, the remaining spectral weights, 8\,eV above the U\,4$f$ core-level main emission lines, is due to satellites arising from configuration interaction.


\subsection{Impact of parameters, $\mu_{\rm dc}$, $U_{\rm ff}$, and $J$, on DFT\,+\,DMFT calculations}	

\begin{figure}[]
	\begin{center}
		\includegraphics[width=0.95\columnwidth]{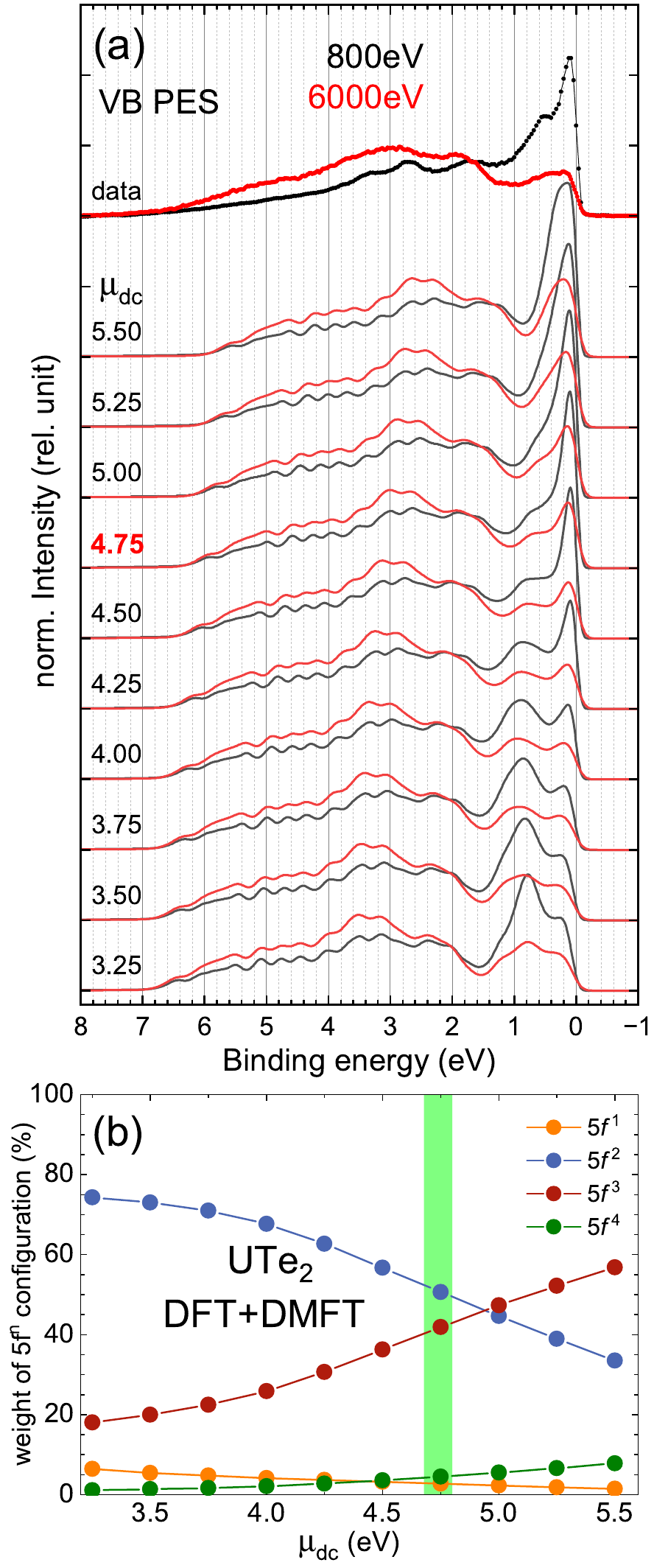}
	\end{center}
	\caption{(a) Experimental VB PES data measured with 800\,eV and 6000\,eV photon energies, compared to DFT\,+\,DMFT spectra calculated with $U_{\rm ff} = 3$\,eV and $J = 0.59$\,eV for several $\mu_{\rm dc}$ values. (b) Variation of the 5$f^n$ weights as a function of $\mu_{\rm dc}$ (in percent). 		The green bar indicates the optimum value of $\mu_{\rm dc}$.}
	\label{LDA_mu}
\end{figure}

Figure\,\ref{LDA_mu}\,(a) displays the $\mu_{\rm dc}$ dependence of the corresponding DFT\,+\,DMFT VB spectra. The spectra are generated by multiplying the orbital-resolved DOS from the DFT\,+\,DMFT calculations with a Fermi function, convoluting with the experimental resolution, and correcting for the respective photoionization cross-sections (see Table I). For comparison, all spectra, both experimental and calculated, are normalized to their integrated area within the displayed energy range. The spectra around $\mu_{\rm dc}$\,=\,4.75\,eV resemble the data best.

The weights of the U 5$f^n$ configurations in the ground state are sensitive to the choice of $\mu_{\rm dc}$, as shown in Fig.~\ref{LDA_mu}\,(b). For values of $\mu_{\rm dc}$ greater than 4.75\,eV (green bar), the 5$f^3$ configuration becomes dominant and also the 5$f^4$ configuration gains weight. For smaller values of $\mu_{\rm dc}$, the 5$f^2$ configuration becomes more prominent, eventually reaching a situation as in UGa$_2$~\cite{Marino2024}. 
This behavior is expected as $\mu_{\rm dc}$ renormalizes the U 5$f$ orbital energy. In a simplified two-configuration picture, the 5$f^2$ configuration lies lowest in energy for small $\mu_{\rm dc}$, whereas for larger values, the 5$f^3$ configuration becomes energetically favored, as the U 5$f$ energy is given by $\varepsilon^{\rm DMFT}_{5f} = \varepsilon^{\rm DFT}_{5f} - \mu_{\rm dc}$ (where $\varepsilon^{\rm DFT}_{5f}$ is the 5$f$ energy in the DFT result). Coherent charge fluctuations can occur between the two configurations, mediated by inter-site hybridization (itinerant, intermediate-valent case). When the energies of the two configurations are quasi-degenerate, on-site fluctuations (local, mixed-valence case) between 5$f^2$ and 5$f^3$ may occur.



\begin{figure}[]
	\begin{center}
		\includegraphics[width=0.98\columnwidth]{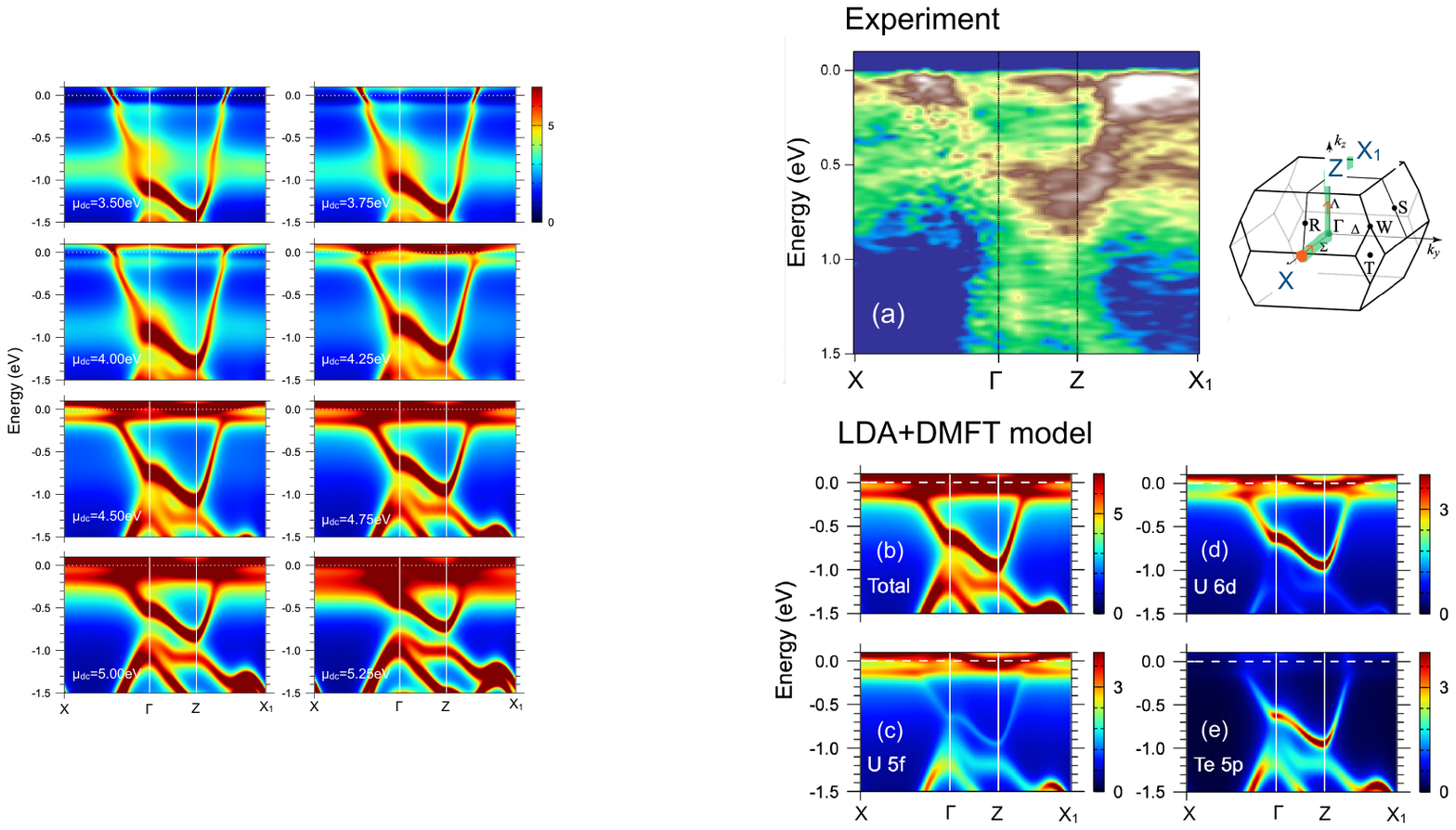}
	\end{center}
	\caption{Momentum-resolved spectral densities in the DFT+DMFT results for selected $\mu_{\rm dc}$ values. The orbital-resolved densities for the optimal $\mu_{\rm dc}$ and the experimental data are shown in Fig.~\ref{ARPES}.}
	\label{ARPES_mu}
\end{figure}

Figure~\ref{ARPES_mu} shows the evolution of the low-energy band structure in the DFT\,+\,DMFT results with varying the double-counting value $\mu_{\rm dc}$. Along the selected path in momentum space, there is a dispersive band extending from $-0.5$~eV at $\Gamma$ to around $-0.8$~eV at the $Z$ point, where its bottom in the adopted path is located. This dispersive band exhibits the hybridized character of the Te\,5$p$ and U\,6$d$ states. The U\,5$f$ states do not contribute in this energy range, as their spectral weight appears near the Fermi level (see panels~(c),(d),(e) in Fig.~\ref{ARPES}). As a result, the dispersive band shifts to higher energies relative to the U\,5$f$ states in a nearly monotonic fashion as $\mu_{\rm dc}$ decreases (corresponding to a shallower 5$f$ level). Thus, a direct comparison of the position of the dispersive band with the ARPES data in Fig.~\ref{ARPES}(a) provides a reasonable way to determine the $\mu_{\rm dc}$ value.

The value $\mu_{\rm dc} = 4.75$~eV yields good agreement in the dispersive feature with the experimental ARPES data. Furthermore, it produces a flat-band feature around $-0.1$ to $-0.2$~eV associated with U 5$f$-dominated states. In addition to the dispersive band, the position and spectral intensities of the U 5$f$ states are also sensitive to $\mu_{\rm dc}$, although their dependence exhibits a more complex behavior due to modulation of the correlation effects arising from changes in the U 5$f$ valency. Therefore, both features show a deviation from the experimental data when $\mu_{\rm dc}$ departs from this optimal value.


\begin{figure}
	\begin{center}
		\includegraphics[width=0.99\columnwidth]{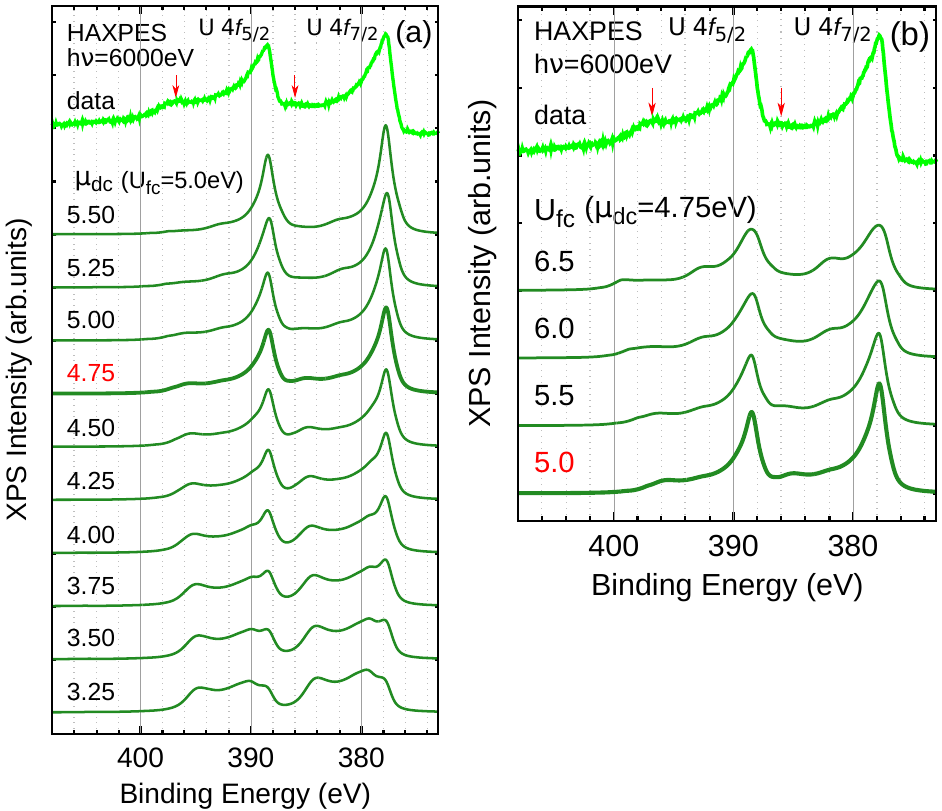}
	\end{center}
	\caption{(a)\,\,$\mu_{\rm dc}$ dependence of DFT\,+\,DMFT\,+\,AIM simulation of U\,4$f$ core-level spectra for fixed $U_{\rm fc}$\,=\,5.0\,eV. (b)\,$U_{\rm fc}$ dependence for fixed $\mu_{\rm dc}$\,=\,4.75\,eV.}  
	\label{corelevel_mu}
\end{figure}

Then we proceed with the core-level spectra. Following Ref.~\cite{Marino2024}, we neglect the orbital degrees of the core state, i.e., $s$-orbital symmetry of the core-orbital is assumed in the simulation to make the computation feasible. In Ref.~\cite{Marino2024}, the multiplet part of the core-valence interaction was found to be unimportant for the 4$f$ core-level spectrum. Furthermore, because the U\,4$f_{5/2}$ and U\,4$f_{7/2}$ levels are sufficiently separated in energy ($\sim 10$\,eV), interference between the two can safely be neglected. Thus, we simulated the total spectrum by accounting for the nominal degeneracy ratio of $6:8$ in the 4$f_{5/2,7/2}$ core states and by applying a 10\,eV shift corresponding to the 4$f$ spin-orbit splitting.

In Fig.\,\ref{corelevel_mu}, U\,$4f$ core-level HAXPES data of UTe$_2$ are compared to the DFT+DMFT+AIM calculations for different $\mu_{\rm dc}$ (in panel (a)) and core-hole potential $U_{\rm fc}$ values (in panel (b)). The best agreement between experiment and theory is achieved for $\mu_{\rm dc}$\,=\,4.75\,eV and $U_{\rm fc}$\,=\,5.0\,eV. As $\mu_{\rm dc}$ decreases below 4.25\,eV, i.e., when enhancing the 5$f^2$ configuration (see Fig.\,\ref{LDA_mu}\,(b)), the satellite becomes at first more pronounced. With further decreasing $\mu_{\rm dc}$ the main emission line develops a substructure, reminiscent of the core-level spectra of strongly localized UPd$_3$. Also for values larger than the optimum $\mu_{\rm dc}$ the  core-level change: the satellite 8\,eV above the main emission line becomes weaker and another satellite, at lower binding energies, appears. The spectral shape also changes with $U_{\rm fc}$. Also here, for large $U_{\rm fc}$, the position of prime satellite changes and  a third contribution appears.


\begin{figure}
	\begin{center}
		\includegraphics[width=0.98\columnwidth]{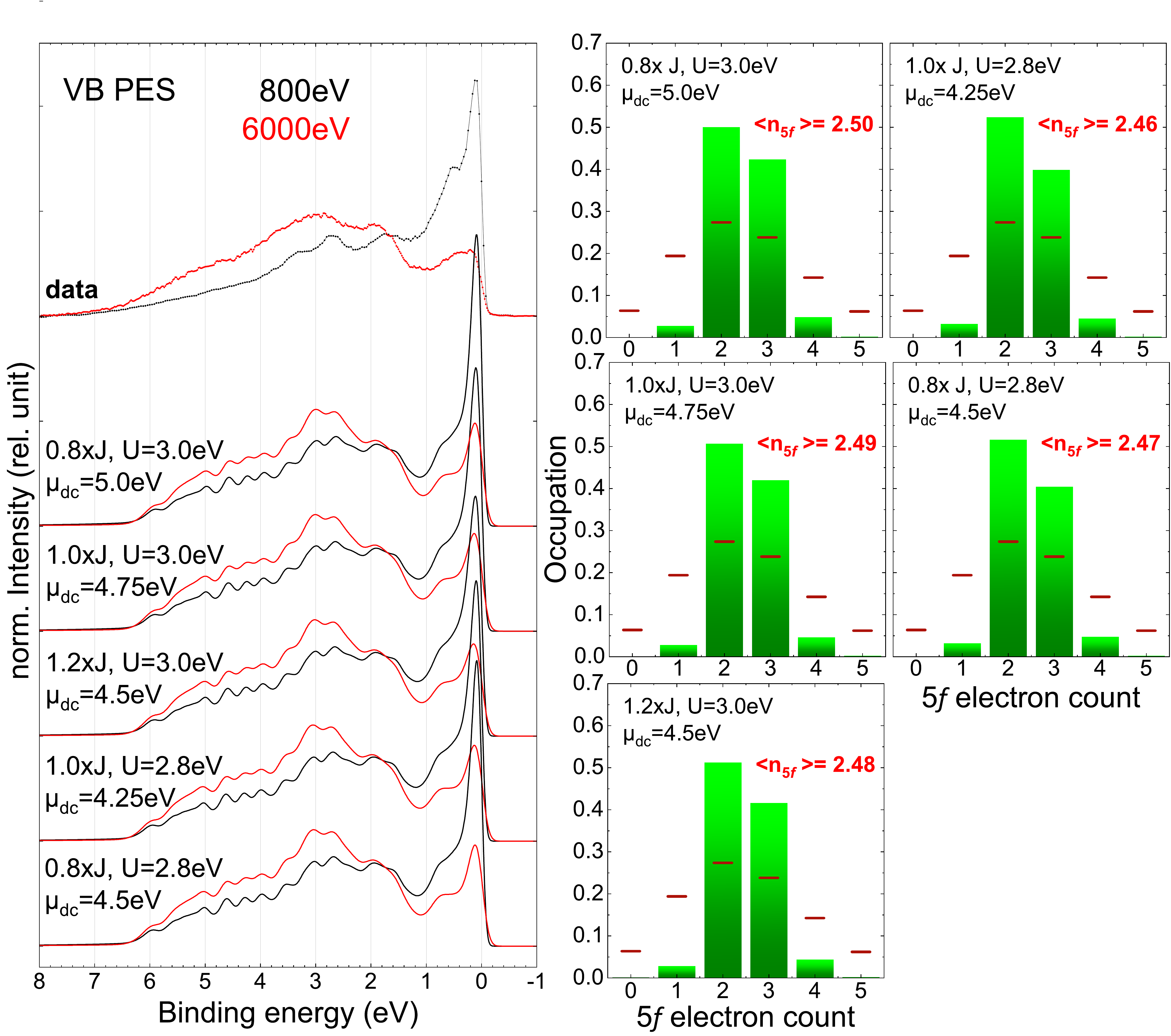}
	\end{center}
	\caption{Left:~Experimental VB PES spectra measured with 800\,eV and 6000\,eV photon energies, compared to DFT\,+\,DMFT spectra for several sets of $U_{\rm ff}$ and $J$ (with $J$ scaled by 0.59\,eV), using adjusted double-counting corrections $\mu_{\rm dc}$ to match the experimental data. Right: Corresponding histograms of the 5$f^n$ weights (in percent). The average 5$f$ occupation $\langle n_{\rm f} \rangle$ remains unchanged across all parameter sets.}
	\label{parameters}
\end{figure}

Finally, we discuss the reliability of our estimation of the U valency with respect to the choice of the Hubbard $U$ and Hund's $J$ parameters. Figure\,\ref{parameters} shows the DFT\,+\,DMFT-calculated VB spectra for several combinations of $U_{\rm ff}$ and $J$ around the values adopted in the above analysis. The double-counting correction $\mu_{\rm dc}$ must be adjusted to ensure agreement between data and calculation. Remarkably, for every parameter set that yields a good match with the data, the weights of the 5$f^n$ configurations in the ground state are found to be nearly identical (see histogram plot in Fig.\,\ref{parameters}). In turn, the experimental spectra are only reproduced when the 5$f^n$ weights are correctly captured.


\subsection{LDA and $\mu_{\rm dc}$ dependence of DFT+DMFT spectral functions for 6$d$-removed modified model}

\begin{figure}[]
	\begin{center}
		\includegraphics[width=0.95\columnwidth]{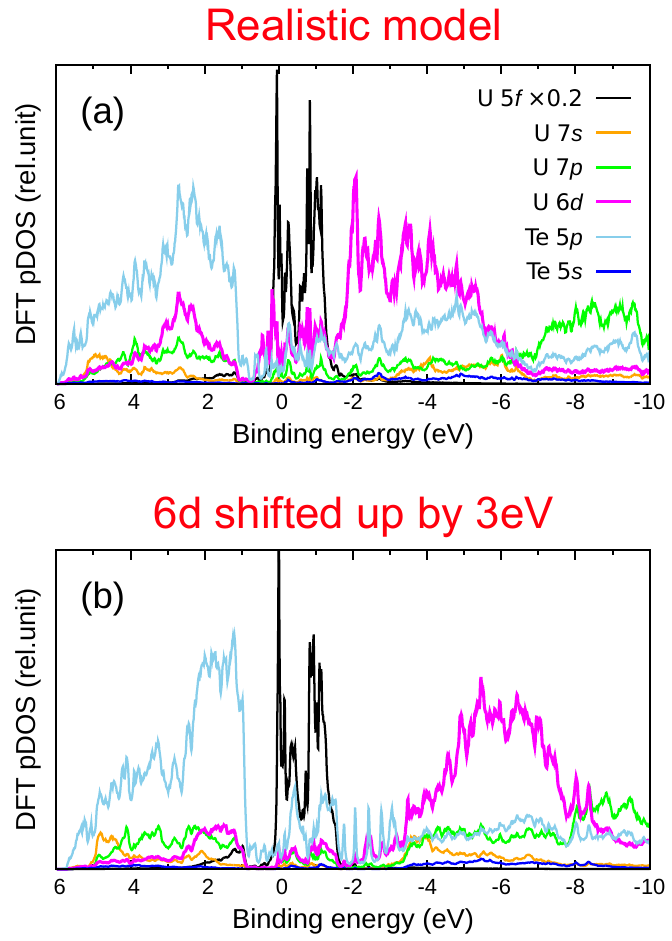}
	\end{center}
	\caption{(a) and (b): DFT electronic structures used as input for the DMFT part in (a) for the \textit{realistic} model, while in (b) for the \textit{modified model}, in which the U 6$d$ states are artificially eliminated from the top region of the VB. }  
	\label{modified_LDA}
\end{figure}

\begin{figure}[]
	\begin{center}
		\includegraphics[width=0.95\columnwidth]{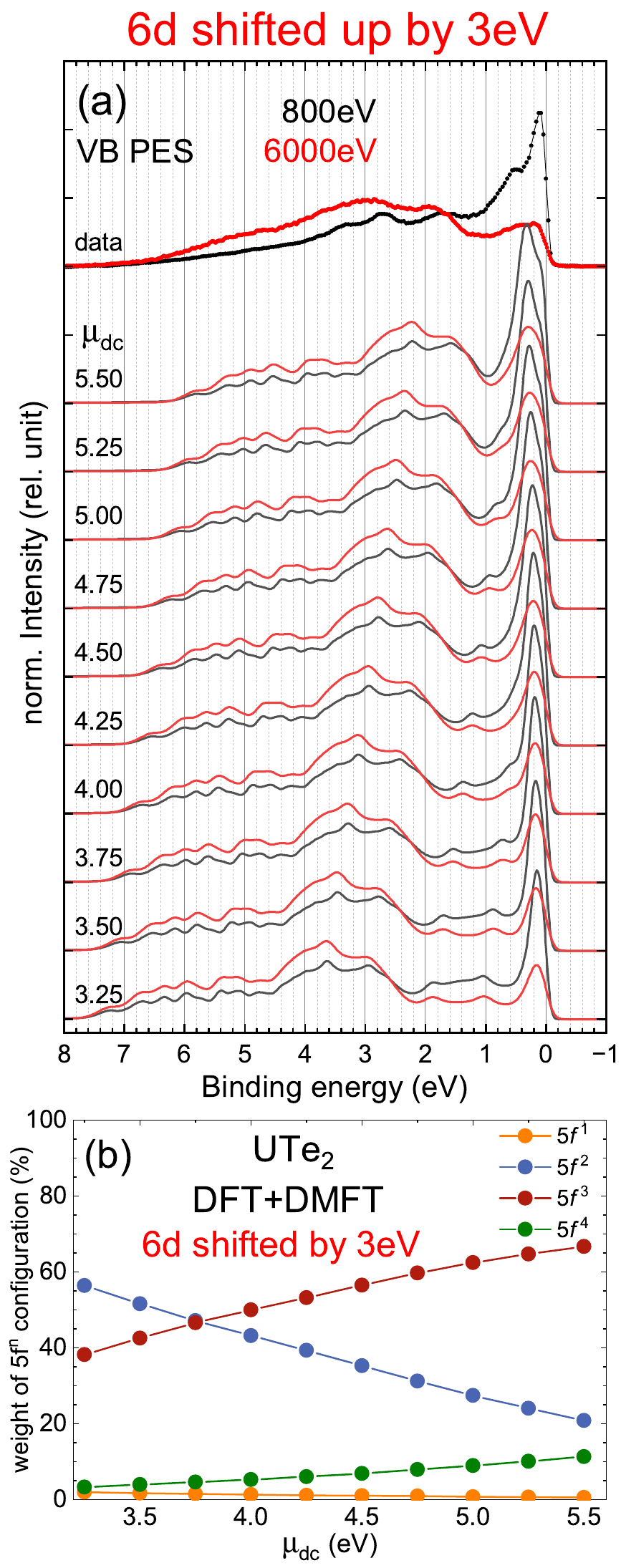}
	\end{center} 
	\caption{(a) Experimental VB PES spectra measured with 800\,eV and 6000\,eV photon energies, compared to DFT\,+\,DMFT spectra calculated for the {\it modified model} with several $\mu_{\rm dc}$ values. The stoichiometric electron count, identical to that of the realistic model, is properly implemented in the DMFT calculations. (b) Variation of the 5$f^n$ weights (in percent) with $\mu_{\rm dc}$ for the modified model.}
	\label{modified_model}
\end{figure}

Figures~\ref{modified_LDA}(a) and (b) show the LDA calculations used as input for the DFT\,+\,DMFT calculations. Panel~(a) corresponds to the realistic model obtained for the experimental crystal structure of UTe$_2$. Practically, the valence spectral densities are computed from the tight-binding Hamiltonian derived from the LDA bands, as described in Sec.~II. Panel~(b), on the other hand, represents the \textit{modified model} in which the participation of the U 6$d$ states in the low-energy valence electronic structure is prohibited. This modified model is obtained by artificially shifting the 6$d$ orbital energies in the LDA tight-binding Hamiltonian upward by 3~eV, which eliminates the U 6$d$ contributions in the region near the top of the VB (in the $-1$ to $0$~eV range). Apart from that, the non-interacting spectra do not differ significantly from the original realistic model (compare Figs.~\ref{LDA_mu}(a) and \ref{LDA_mu}(b)). However, once correlation effects are included within DMFT, this elimination leads to a large impact on the VB features.

The $\mu_{\rm dc}$ dependence of the DFT\,+\,DMFT VB spectra in the modified model is displayed in Fig.~\ref{modified_model}\,(a). Also here the spectra are generated by multiplying the orbital-resolved DOS from the DFT\,+\,DMFT calculations with a Fermi function, convoluting with the experimental resolution, and correcting for the respective photoionization cross-sections (see Table I). For comparison, all spectra, both experimental and calculated, are normalized to their integrated area within the displayed energy range. For the modified model, the VB data close to the Fermi energy cannot be reproduced for any set of parameters, confirming that it is unrealistic and that the U 6$d$ states must be taken into account when modeling the electronic structure of UTe$_2$.

Figure~\ref{modified_model}(b) shows the evolution of the U 5$f^n$ occupation as function of $\mu_{\rm dc}$ in the modified model.    As the 6$d$ states are not accessible, the 5$f^3$ configuration becomes preferably occupied. This suggests U ions favor the U$^{3+}$ configuration in this case. Therefore, the experimental observation of a predominant U\,5$f^2$ configuration in VB-RIXS\,\cite{Miao2020,Christovam2024} strongly implies partial filling of the 6$d$ states. This filling facilitates the U$^{3+}$ character, so that the structure of UTe$_2$ does not have to accommodate two large Te$^{2-}$ ions. Instead, one of the two Te ions can adopt the smaller configuration of Te$^{1-}$.

%

\end{document}